\documentclass{aa}

\usepackage{graphicx}

\usepackage{txfonts}

\begin{document}

\title{Tides in asynchronous  binary systems}

\author{O. Toledano\inst{1}, E. Moreno\inst{2}, G. Koenigsberger\inst{1}, R. Detmers\inst{3}, 
   \and
 N. Langer\inst{3}
       }

\offprints{G. Koenigsberger}

\institute{Instituto de Ciencias F\'{\i}sicas UNAM, Apdo. Postal 48-3, Cuernavaca, Mor. 62251\\
           \email{oswaldo@ce.fis.unam.mx; gloria@ce.fis.unam.mx}
       \and
          Instituto de Astronom\'{\i}a, UNAM, Apdo. Postal 70-264, Mexico D.F. 04510 \\
           \email{edmundo@astroscu.unam.mx}
      \and
          Sterrenkundig Instituut, Universiteit Utrecht, Postbus 80.000, Utrecht, The Netherlands  \\
           \email{R.G.Detmers@astro.uu.nl; N.Langer@astro.uu.nl}
        } 

\date{Received   :  }

\abstract
{Stellar oscillations are excited in non-synchronously rotating stars in binary systems due to the 
tidal forces.   Tangential components of the tides can drive a shear flow which behaves as a 
differentially forced rotating structure in a stratified outer medium. 
}
{The aims of this paper are to show that our single-layer approximation for the calculation of the forced 
oscillations yields results that are consistent with the predictions
for the synchronization timescales in circular orbits, $\tau_{sync}\sim a^6$, 
thus providing a simplified  means of computing the energy dissipation rates, $\dot{E}$.
Furthermore,  by calibrating our model results to fit the relationship between synchronization
timescales and orbital separation, we are able to constrain the value of the kinematical viscosity 
parameter, $\nu$.  
}
{We compute the values of $\dot{E}$ for a set of 5 M$_\odot+$4 M$_\odot$ model binary systems 
with different orbital separations, $a$, and use these to estimate the synchronization timescales.}
{The resulting $\tau_{synch}$ {{\it vs.}} $a$ relation is comparable to that of Zahn (1977) for
convective envelopes, providing a calibration method for the values of $\nu$.  For the 
4$+$5 M$_\odot$  binary modeled in this paper, $\nu$ is in the range 0.0015 -- 0.0043 R$_\odot ^2$ /day for 
orbital periods in the range 2.5 -- 25 d.  In addition,  $\dot{E}$ is found to decrease by $\sim$2 orders of
magnitude as synchronization is approached, implying that binary systems may approach
synchronization relatively quickly but that it takes a much longer timescale to actually
attain this condition.}
{The relevance of these results is threefold: 1) our model allows an estimate for the numerical 
value of $\nu$ under arbitrary conditions in the binary system;  2) it can be used to calculate 
the energy dissipation rates throughout the orbital cycle for any value of eccentricity and 
stellar rotational velocity; and 3) it provides values of the tangential component of the 
velocity perturbation at any time throughout the orbit and predicts the location on the 
stellar surface where the largest shear instabilities may be occurring.  
We suggest that one of the possible implication of the asymmetric distribution of $\dot{E}$ 
over the stellar surface is the generation of localized regions of enhanced surface activity.}

\authorrunning{Toledano et al.}
\titlerunning{Asynchronous binary systems}
\maketitle

\section{Introduction} 

A binary system is said to be in equilibrium when the orbit is circular, the stellar rotation period 
and the orbital period are synchronized and the spin axes of both stars are  aligned and perpendicular
to the orbital plane. In any other case, oscillations are excited and the tidal deformations which 
are time-dependent become inevitably dissipative. We will focus on the situation in which the stellar 
rotation period differs from the orbital period, that is, a non-synchronously rotating system.
To describe the degree of non-synchronicity, we define $\beta=\omega_0/\Omega$, where $\omega_0$ and $\Omega$
are the rotational and orbital angular velocities, respectively.

As in the case of the terrestrial tides, the resulting asymmetrical shape of the star is due 
to a second order gravitational effect, the differential force. Thus, the  rising tidal bulges, 
opposite to each other and comparable in size, are  associated with the so called, 
equilibrium tide. Additionally,  a  set of surface oscillations commonly
known as the dynamical tide,  characterized by a wide variety of harmonic frequencies, is established.  
This oscillatory response of the star depends on $\beta$ and on  the stellar and orbital parameters. 
In addition, it depends on the capability of the stellar material to transfer angular momentum 
and energy between its different layers, which is represented by the viscosity, $\nu$. 

The tidally induced oscillations have both  radial and  azimuthal components. If we focus on the
azimuthal component and keep in mind that the tidal forces decrease as r$^3$, with $r$ the 
distance between a mass element and the center of the  perturbing star, it is evident 
that since an external layer of a perturbed star is subjected to a larger tidal force than a layer below 
it, a differentially rotating structure will be produced in any $\beta\neq$1 binary system. 
If $\nu\neq$0,  the interaction between the different layers takes the form of a shearing flow,
thereby leading to energy dissipation.  The viscosity and the azimuthal velocity are 
the crucial parameters defining the amount of energy that is dissipated, $\dot{E}$.   

The long term trend of the angular momentum transfer and dissipative effects  in binary systems 
is one in which the rotation frequency becomes synchronous with the orbital frequency.  
The theoretical framework for the calculation of tidal synchronization was developed
by Zahn (1966, 1977, 1989), who invoked the tidal friction as a physical mechanism asociated with 
tidally excited oscillations which contribute to the  angular momentum transfer. In stars with 
convective envelopes, redistribution of angular momentum occurs through dissipation processes like 
turbulent viscosity, retarding the equilibrium tide.  In stars with radiative envelopes, the 
mechanism found by Zahn (1975, 1977) is the action of radiative damping taking place on the dynamical 
tide.  Zahn (1977, 1989) found
the characteristic timescale for synchronization as a function of $a$, the major semi-axis 
of the orbit;  $\tau_{sync}\sim a^6$ for stars with convective envelopes and $\tau_{sync}\sim a^{8.5}$ 
for stars with radiative envelopes.

On orbital timescales, the tidal oscillations can produce observable photometric and absorption 
line-profile variability (see, for example, Smith 1977; Gies \& Kullavanijaya 1988; Willems \& Aerts 2002).
Moreno \& Koenigsberger (1999) presented a simplified model for the calculation of the 
surface oscillations of a star driven by the tidal interaction in a binary system.  The
calculation is performed for a thin surface layer of small surface elements distributed
along the equatorial belt of the star.  The solution of the equations of motion for these
surface elements provides the velocity perturbations in the radial and azimuthal directions
as a function of time within the orbital cycle of the binary. In the non-inertial
reference frame used to solve the equations of motion the model  includes the gravitational forces 
of both stars,  gas pressure, centrifugal, coriolis, and viscous forces.   In the model we
compute only the motion of the  center of mass  of each surface element, thus ignoring the
details of the inner motions within these elements.  In this approximation, the buoyancy
force is not taken into account.
Finally, we note that the results in this paper are obtained using a polytropic
index n=1.5, corresponding to a rigorous convective envelope, instead of n=1 as was used 
previously (Moreno et al. 2005; Koenigsberger et al. 2006), although the difference between the 
results obtained for these two values of n is not significant.

Moreno et al. (2005) extended the model to include the computation of the photospheric absorption 
lines that would be produced by the perturbed stellar surface.  This {\it ab initio} calculation 
of the line profile variability due to the forced tidal oscillations was applied to the case of
$\epsilon$ Per, showing that from a qualitative standpoint, the predicted line profile
variability is comparable to the observations. 
More recently, the model was used to analyze the
peculiar line-profile variability in the optical counterpart of the X-ray binary 2S0114$+$650
(Koenigsberger et al. 2006), leading to the conclusion that the tidal perturbations could be
the ``trigger" for the periodic ejection of wind material at densities that are higher than
the average wind density.  They hypothesize that these effects result from the combined action
of the shear energy dissipation, the radial oscillations and the radiation pressure.   The 
significance of this result is that, in addition to providing a tool for studying line profile 
variability in binary systems, the model provides a means of constraining the value of the viscosity,
which is the parameter involved in the energy dissipation mechanisms. Thus, although the model
is designed to study the variability in binary systems on  orbital timescales, it is of interest
to explore whether it yields results that are consistent with the long-term behavior,
specifically, the tidal evolutionary theories.  That is, using the value of viscosity and the 
angular velocity perturbations  obtained from the calculation, an estimate of the rate of dissipated 
mechanical energy, $\dot{E}$, can be derived.  The timescale over which $\dot{E}$ becomes negligible 
is a measure of the synchronization timescale  of the binary system.  
Thus, the results of this one-layer calculation of the tidal oscillations may be compared with 
the predictions of tidal synchronization theory such as those  of Zahn (1977, 1989) yielding, 
on the one hand, a test of the Moreno et al. (2005) model and, on the other, a  constraint on the 
numerical values of the viscosity.  These are two of the objectives of this paper.

The expression for the calculation of the viscous energy dissipation rate is derived in
Sections 2 and 3; in Section 4 we present the results of our model calculations for a
5 M$_\odot$$+$ 4 M$_\odot$ binary system at different orbital separations and compare our predictions
with the synchronization timescales of Zahn;  and in Section 5 we discuss the implications of
this result  and present the conclusions.


\section{Viscous angular momentum transport in a shearing medium}

The approximate treatment of angular momentum transfer between the equatorial surface layer and the 
rigidly-rotating inner region of the star is made assuming that the equatorial external layer behaves as 
a shearing thin disk, in analogy with the dissipation mechanisms described for accretion disks 
(see for example, Frank et al. 1985,  Lynden-Bell and  Pringle, 1974). In the assumed differentially 
rotating medium, tangential stresses between adjacent layers are assumed to be the mechanism of 
transport of angular momentum.  Included in these is the angular momentum transport due to magnetic 
torques, which has been shown to be the dominant mechanism in recent stellar model
calculations (Petrovic et al. 2005) where the formulation of Spruit (2002) is used to treat the
magnetic field.   In our calculations, the mechanisms by which transfer of angular momentum in the 
shear flow occurs is  englobed in the viscosity.



We adopt the approach for the transport of angular momentum  that
consists of imagining an idealized situation in which parts of a ring-shaped outer layer 
slide over a rigidly-rotating inner region. If friction between adjacent layers is assumed
to exist and if the angular velocity increases or decreases outwards, a net torque will be 
exerted by the outer ring on the inner region. The resulting torques will work towards synchronizing 
the outer and inner region, dissipating energy in the process. 

For this mechanism to be active, the particles in the external ring have to be able to interact 
with those in the inner rotating region.  However, contrary to the case of many accretion disk
models where the particle velocity is Keplerian, the velocity of different layers on a stellar 
surface is determined by the stellar rotation, the tidal forcing and the viscosity 
of the material.

Let us  assume that the viscosity in the gaseous shearing medium is a turbulent one. Gas elements 
with  small random motions at a typical turbulent velocity $v_{turb}$, travel a  distance 
$\lambda_{turb}$ (the characteristic length and velocity scales of the turbulence), in all directions 
before mixing with the surroundings. There exists a typical turbulent viscosity,

\begin{equation}
\nu \sim \lambda_{turb} v_{turb}
\end{equation}  

However, both $\lambda_{turb}$ and $v_{turb}$ are unknown parameters. A clever solution to this problem 
was proposed in the early 1970' s by Shakura and Sunyaev (1973), consisting of a simple parametrization of 
the two turbulence scales.   For the length scale, they assumed isotropy 
and characterized it by the typical size of the largest turbulent eddies which cannot exceed the 
thickness $H$ of the assumed ring $\lambda_{turb} < H$. For the velocity scale, a more straightforward 
scaling could be made. That is, if the turbulent velocity were supersonic, emerging shock waves would 
dissipate the energy and reduce the velocity to that of, or below, the sound speed. This leads to the 
choice of a $v_{turb}$ that is  subsonic, thus  $v_{turb} < C_{s}$,  and the viscosity can be written; 

\begin{equation}
\nu \sim \alpha H C_{s} 
\end{equation}  

\noindent with  $\alpha < 1$. This is the well-known $\alpha$-prescription of 
Shakura \& Sunyaev (1973).

The physical idea of the viscous torque $g$ is to assume  that, if the gas flow is turbulent to 
some extent, gas particles of adjacent layers will be exchanged  in the radial direction (Frank 1985). 
Since the two different radial flows of material originating in both layers will have  different specific angular momenta, this will cause a net transfer of angular momentum between them. 

We can then calculate the magnitude of the viscous torque on the equatorial  stellar region
in terms of a kinematic turbulent viscosity $\nu$, which  can be expressed as $ \nu = \mu/\rho$. 
Here $\rho$ is the local density of the gas and $\mu$ is the coefficient of shear viscosity 
which is related to the  component ${\tau}_{r\phi}$ of the viscous stress tensor in cylindrical 
coordinates (r, $\phi$, z) by (Landau and Lifshitz 1984):

\begin{equation}
\tau_{r \phi} = \mu ( \frac{1}{r}\frac{\partial u_r}{\partial \phi} + \frac{\partial u_\phi}{\partial r} - \frac{u_{\phi}}{r}  )
\end{equation}

\noindent where $u_r$, $u_\phi$ are the radial and azimuthal components of the velocity field. 
Now, inserting $u_\phi = \omega_{\phi} r$ with $\omega_{\phi}$ as the rotational angular 
velocity of the tidal shear flow, 

\begin{equation}
\frac{\partial u_\phi}{\partial r} = \frac{\partial (r\omega_{\phi})}{\partial r} = \omega_{\phi} + r \frac{\partial \omega_{\phi}}{\partial r},
\end{equation}  

\noindent and  neglecting the shear effects in the radial direction, the viscous stress tensor becomes, 

\begin{equation}
\tau_{r \phi} = \mu r \frac{\partial \omega_{\phi}}{\partial r}
\end{equation}  

This is a tangential force per unit area, and multiplying  $\tau_{r \phi}$ by the lever arm $r$, 
gives the viscous torque per unit area exerted by the sliding ring-shaped outer layer on the 
rigidly-rotating inner region.  Integrating over the area of interaction ($dA = r d\phi dz$),

\begin{equation}
g =  \int r\tau_{r\phi} dA  
=  \int  \mu r^3 \frac{\partial \omega_{\phi}}{\partial r} d\phi dz
\end{equation}  

\begin{equation}
=  \nu  \int\limits^{2\pi}_0 r^3 \frac{\partial \omega_{\phi}}{\partial r}  d\phi \int\limits^{H/2}_{-H/2} \rho dz   \hspace{2 mm}   
\end{equation}  

\begin{equation}
=   \nu \Sigma \int\limits^{2\pi}_0  r^3  \frac{\partial \omega_{\phi}}{\partial r} d\phi
\end{equation}  

\noindent where we have used that the density per unit area on the equatorial plane of the
equatorial stellar region is  

\begin{equation}
\Sigma = \int\limits_{-H/2}^{H/2} \rho dz
\end{equation}  

\noindent with $H$ the width of the equatorial stellar region.  With 
   $\Sigma = {\rho}H$,  the torque can also be expressed as
\begin{equation}
g =  \nu \rho H \int\limits_0^{2\pi} r^3  \frac{\partial \omega_{\phi}}{\partial r} d\phi
\end{equation}  

According to this model, no net angular momentum is transported unless there is shearing differential 
rotation; that is, the condition for net angular momentum transport is, 
$\partial \omega_{\phi}/\partial r \neq 0$.  



\section{The energy dissipation caused by the differential viscous torque}

A natural consequence of the shearing differential rotation is the dissipation
of energy, $\dot{E}$. Let us consider the energy dissipation caused by the viscous torque
acting on the stellar equatorial external layer between $r$ and $r$ + ${\Delta}r$.
The net differential viscous torque on this layer is 

\begin{equation}
g(r+\Delta r) - g(r)\sim \frac{\partial g}{\partial r} \Delta r 
\end{equation}  

Since the torque is acting in the direction of the angular velocity $\omega_{\phi}$, the rate of work exerted onto the external layer is;

\begin{equation}
\omega_{\phi} \frac{\partial g}{\partial r} \Delta r =  \frac{\partial }{\partial r} (g\omega_{\phi}) \Delta r- g \frac{\partial \omega_{\phi}}{\partial r} \Delta r  
\end{equation}  

\noindent where the first term of the right-hand side represents  the rate of transported mechanical 
energy through the gas by the torque, and the term $-g (\partial \omega_{\phi}/\partial r)\Delta r$, 
the corresponding local rate of loss of mechanical energy converted into heat in the gas. We are 
interested in the dissipation caused by the viscous torques within the gas at a rate 
$g (\partial \omega_{\phi}/\partial r)\Delta r$ on the external layer of width $\Delta r$. 

The local dissipation rate on an equatorial surface element with an azimuthal width ${\Delta}\phi$ is

\begin{equation}
\dot{E}_{elem} = g \frac{\partial \omega_{\phi}}{\partial r} \Delta r = \nu \rho H r^{3 }  \left \{  \frac{\partial \omega_{\phi}}{\partial r} \right \}^{2} \Delta r  \Delta \phi 
\end{equation}  

\begin{equation}
 = \nu \rho  \left \{  r \frac{\partial \omega_{\phi}}{\partial r} \right \}^{2} r  \Delta \phi \Delta r H 
\end{equation}  

Since we require an estimate of the dissipation rate on the entire surface of the star, the previous 
description must be  extended to include the dissipation produced at polar angles ($\theta$). 
An approximate  generalization can be performed considering that the same dissipation mecanisms 
apply to a series of rings of thickness $r \Delta \theta$, situated across the star from the north 
pole to the south pole, and the elements in these rings with an azimuthal angle $\phi$  have the same 
angular velocity as that of the equatorial element with the same azimuth $\phi$. This approximation 
was used by Moreno et al. (2005) for the line-profile calculation.


From Eq. (14), the local dissipation rate on any surface element is

\begin{equation}
\dot{E}_{elem}  = \nu \rho \left \{  r  \frac{\partial \omega_{\phi}}{\partial r} sin \theta \right \}^{2} r (sin \theta) \Delta \phi \Delta r (r \Delta \theta)
\end{equation}  

\noindent with ${\omega}_{\phi}$ the angular velocity of the equatorial
   surface element in the meridian of the element. Integrating in polar
   angle, we obtain the dissipation rate on a meridional shell,

\begin{equation}
\dot{E}_{merid.shell}  = \nu \rho \left \{  r  \frac{\partial \omega_{\phi}}{\partial r} \right \}^{2} r^{2} \Delta \phi \Delta r \int\limits_{0}^{\pi}  sin^{3}\theta d\theta
\end{equation}  

\begin{equation}
 = \frac{4}{3}\nu \rho \left \{  r  \frac{\partial \omega_{\phi}}{\partial r} \right \}^{2} r^{2} \Delta \phi \Delta r 
\end{equation}  
To obtain the total dissipation rate ${\dot{E}}$, this equation
   must be integrated over the azimuthal coordinate $\phi$. We use
   the following approximation:
$(\partial \omega_{\phi}/\partial r) \sim (\Delta \omega_{\phi}/ \Delta r)_{i}$, where 
 $\Delta \omega_{\phi} \sim \Delta V(\phi_{i})/ r_{i}$, with $r_i$ the radial position of 
 the $i$-th equatorial
   surface element, and ${\Delta}V({\phi}_i)$ the difference of
   azimuthal velocity of the element and the inner stellar region.
   Then, with $({\Delta}r)_i$ the radial thickness of the element,
   the total energy dissipation can be obtained as a sum over the $n$
   equatorial surface elements,

\begin{equation}
\dot{E} \sim  \frac{4}{3}\nu \rho  \sum_{i}^n  \left \{  r^{2}\frac{\Delta \omega_{\phi}}{\Delta r}\right \} _{i}^{2} \Delta \phi_{i} (\Delta r)_{i}  
\end{equation}  

\begin{equation}
 \sim \frac{4}{3} \nu \rho  \sum_{i}^n \left \{ \frac{r_{i} l_{i} }{(\Delta r)_{i}}  (\Delta V(\phi_{i}))^{2} \right \} 
\end{equation}  


\noindent where additionally, $ r_{i} \Delta \phi_{i}$ is simplified by the factor $l_{i}$, which 
represents the azimuthal length of the element.  

\section{Synchronization timescales}

For a super-synchronously rotating star, 
the transfer of angular momentum tends to brake the outer stellar layers, causing the total energy of
the system to gradually decline. This, in turn, drives the system towards  synchronization and,
therefore, $|\Delta V(\phi_{i})| \rightarrow 0$.  

Zahn (1977, 1989) found two different relations between the synchronization timescale and the 
orbital separation in binary systems, depending on whether the star's outer envelope is radiative, 
$\tau_{Zahn}\sim(a$/R$_1$)$^{8.5}$, or convective, $\tau_{Zahn}\sim(a$/R$_1$)$^6$.  He  proposed  that for 
stars with convective  envelopes, the most efficient synchronizing mechanism is due to turbulent viscosity. 
In what follows, we will concentrate on this formulation, a choice that is  motivated 
{\it a posteriori} by the results we derive and by the possibility that,
even in stars with radiative envelopes, turbulence is induced by the differential rotation.

Zahn (1977) gives in his Eq. 6.1 an approximate characteristic  synchronization
timescale  

\begin{equation}
\tau_{Zahn} \sim q^{-2} (a/R_{1})^6 \sim 10^{4}((1+q)/2q)^{2}P^{4} \hspace{1.0cm}  years
\end{equation}

\noindent where $q=M_{2}/M_{1}$ is the mass ratio of the secondary to the primary star and $a$
is the orbital separation in solar units, R$_1$ is the equilibrium radius of the star in solar
units and P is the orbital period in days.

Adopting our  one-layer model for  calculating  the tidal oscillations, and assuming that
all of the energy dissipation associated with the shearing forces occurs on this surface layer,
we can estimate  the timescale required for the rotation period of the star to be 
synchronized with the orbital period of its companion as follows.
If E$^{init}_{rot}$ and E$^{sync}_{rot}$ are the present and synchronization stellar rotational energies,
computed with the moment of inertia of a rigid sphere and the azimuthal velocity at the stellar equator, 
E$_{rot} \sim$(1/2)I($\omega)^2\sim$ (1/5)M$_1$ V$_{rot}^2$, then a rough estimate of the timescale 
needed to  synchronize the system is
  
\begin{equation}
\tau_{sync} \sim \frac{\frac{1}{5}M_{1} (V_{rot}^{init})^{2}-\frac{1}{5}M_{1} (V_{rot}^{sync})^{2}}{\frac{4}{3} \nu \rho  \sum_{i}^n \left \{ \frac{r_{i} l_{i} }{(\Delta r)_{i}}  (\Delta V(\phi_{i}))^{2} \right \} }
\end{equation}  



 
\subsection{Comparison with Zahn's synchronization timescales}

In order to compare the above synchronization rates with those that are
inferred from Zahn's ($a$/R$_1$)$^6$ prediction, our code was used to compute  
values of $\Delta V(\phi_{i})$ and r$_i$ for a set of  binary models with 
different orbital periods. 

The test binary system that was selected consists of a Main Sequence intermediate-mass
system with $M_{1}=$5 M$_{\odot}$, M$_{2}=$4 M$_{\odot}$ and R$_{1}=$3.2 R$_{\odot}$ in a
circular orbit. Observational evidence (Tassoul 2000;  Pan 1997;  Mathews \& Mathieu 1992; 
Witte \& Savonije 1999) provides constraints for the circularization timescales of such systems,
and   $\tau_{sync}$  is found to be  shorter than $\tau_{circ}$ (Claret \& Cunha 1997; 
Zahn 1977; Hut 1981).   

We chose to fix the value of $\beta= \omega_{0}/\Omega=$2, which means that  for decreasing
orbital separations, the rotational angular velocity, $\omega_{0}$, increases proportionately with
the increasing  orbital angular velocity $\Omega$.  Although this might seem like a somewhat arbitrary
choice, it is important to note that it is $\beta$ rather than V$_{rot}$ that plays a fundamental role in
defining the frequencies and amplitudes of the oscillations.  An intuitive way this can be understood
is to consider two binary systems  with different $\omega_0$, but both with $\beta=$1.  Because 
$\beta=1$ corresponds to the equilibrium condition, no oscillations are present, regardless of 
how much larger one of the values of $\omega_0$ may be with respect to the other one.

\begin{figure}
\centering
\includegraphics[width=9cm]{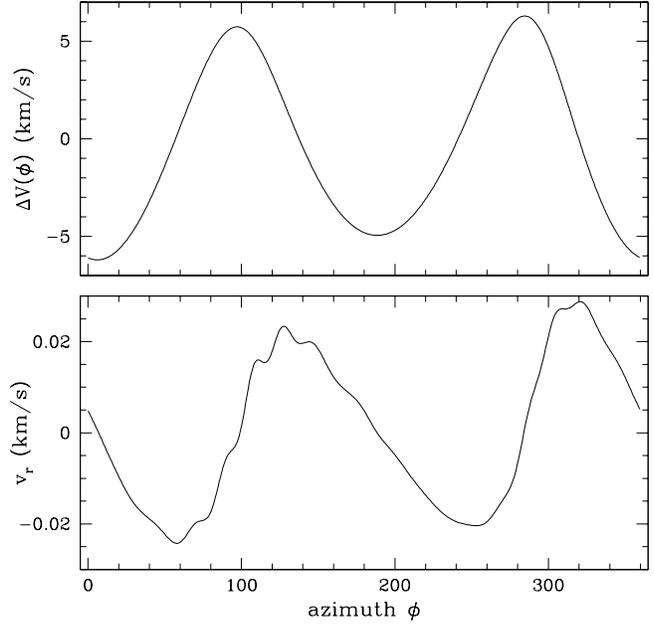}
\caption{Top: The difference of azimuthal velocity  of each surface element and the rigidly
rotating inner stellar body, $\Delta$V($\phi$) along  the equatorial region for
a P$_{orb}=$5 d binary system.  The angle $\phi$ is measured with respect to the line
connecting the two stars.   Bottom:  The component of the velocity in the radial direction 
of each surface element along the equatorial belt for the same model. The star is rotating 
at twice the orbital frequency.
}
\end{figure}

Figure 1  illustrates the  example of  $\Delta$V($\phi$) and V$_r$
along the equatorial belt  for the P$_{orb}=$5 day  binary system.  The abscissa
corresponds to the location of the surface element along the equator with respect to the line
connecting the two stars. The largest  differences between the surface azimuthal velocity
and the inner rigid-body rotating region are in general associated with the tidal bulges
($\phi\sim$0 and 180$^\circ$) and the locations perpendicular to the axis of the orbit, 
and it is at these locations where the largest energy dissipation occurs. Note that the
tangential component is $\sim$2 orders of magnitude larger than the radial component, making
it the dominant source of energy dissipation.\footnote{It is also the dominant source of
photospheric absorption line profile variability as shown by Moreno et al. 2005.}  
When the orbital separation
decreases, the azimuthal variation is no longer smooth, with shorter scale oscillations appearing,
as illustrated in Figure 2 where we plot $\Delta$V($\phi$) for the P$_{orb}=$2.5 day  binary system. 
Because we are considering circular orbits, $\Delta$V($\phi$)  shown in Figures 1 and 2 do not 
change significantly as a function of orbital phase and it is possible to assign a 
single $\Delta$V($\phi$) curve to each of the models.  This would not be the case if the orbit 
were eccentric, where strong changes occur between periastron and apastron phases.

\begin{figure}
\centering
\includegraphics[width=9cm]{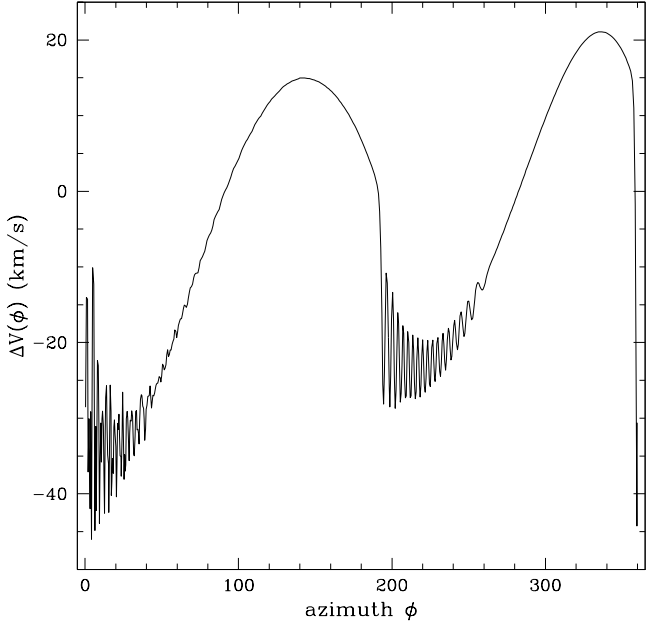}
\caption{The azimuthal velocity  $\Delta$V($\phi$) along the equatorial region of surface
elements with angle $\phi$  at a given time, for the P$_{orb}=$2.5 day binary system and with
$\delta$R$=$0.01 R$_1$ and $\nu=$0.0043 R$_\odot^2$ d$^{-1}$.
}
\end{figure}

The value of the external oscillating layer's density, $\rho$, was obtained by computing 
stellar structure models using the  code described in Yoon \& Langer (2005). The  value  
$\rho=$10$^{-7}$ g cm$^{-3}$ that we adopted  corresponds to 
the average density in the external 73 layers of the theoretical 5 M$_\odot$ star.  These
73 layers comprise approximately $\delta$R/R$_1=$0.01, which is the thickness of the outer 
layer that we adopted for the oscillations calculation.

The remaining input  parameter that needs to be defined for the oscillations code as 
well as for computing the synchronization times using Eq. (21) is the viscosity, $\nu$.   
Our practice (see Moreno et al. 2005) has been to use the smallest value of $\nu$
allowed by the code.  That is, when the oscillation amplitudes are very large, the
surface elements may either overlap with or become detached from their neighboring elements,
at which time the computation can no longer proceed.  This is prevented by assigning
a sufficiently large value to the viscosity parameter.  Our initial set of calculations
were performed with $\nu=$0.005 R$_{\odot}^2$ day$^{-1}$, which is slightly larger than
the minimum $\nu$ allowed for the P$_{orb}=$2.5 day binary system.  The results yielded  
a least-means-square fit with $\tau_{sync}$ $\alpha$($a$/R$_1$)$^{5.7}$, which is similar 
to that of $\tau_{Zahn}$ $\alpha$ ($a$/R$_1$)$^6$, but displaced vertically on the 
log$\tau_{sync}$ {\it vs} $a$ plane. Thus, the next step consisted in adjusting 
the values of $\nu$ to make $\tau_{sync}$ coincide with $\tau_{Zahn}$.  The resulting values 
of $\nu$ and energy dissipation rates are listed in the last two columns of Table 1, and the 
synchronization timescales are plotted in Figure 3.  Note that the 0.0015--0.0043 R$_\odot^2$/d 
range in $\nu$ suggests that part of the turbulent viscosity is itself produced through the 
tidal perturbations processes, since the closest binaries require the largest values of $\nu$.  

\begin{figure}
\centering
\includegraphics[width=9cm]{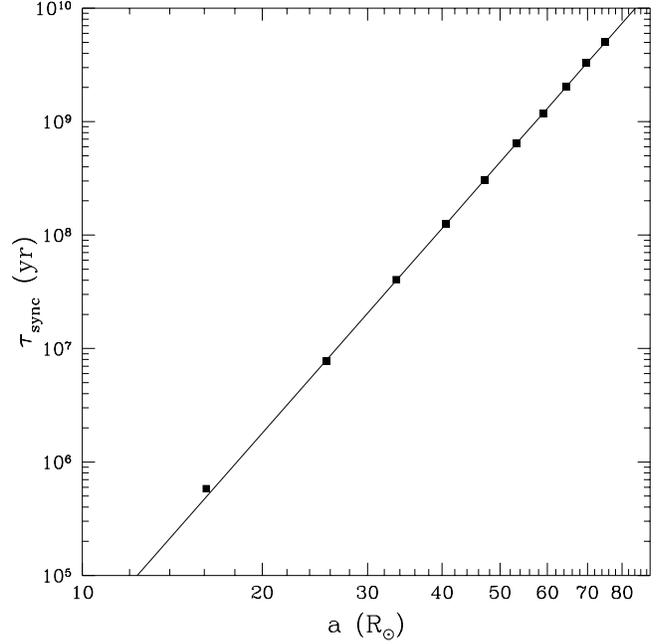}
\caption{Synchronization timescales (squares) calculated from our tidal oscillation model 
for a fixed $\beta=$2, and $\delta$R$=$0.01 R$_1$, and for values of $\nu$ as listed in Table 1 
compared with  Zahn's (1977) relation for stars with convective envelopes (continuous line). 
}
\end{figure}

We thus find that the shear energy dissipation due to the tangential velocity components 
of the surface layer used in our approximate treatment of the tidal interaction leads to
synchronization timescales that are consistent with those predicted by Zahn's  
treatment of the star's entire structure for the case in which the outer envelope is
convective.   

 


\begin{table}[htdp]
\caption{Binary system models for the $\tau_{sync}$ {\it vs.} $a$  calculation.}
\centering
\begin{tabular}{cccccc}
\hline\hline
P$_{orb}$  &  $a$   &  V$_{orb}$  &  V$_{rot}$ & $\nu$  &  $\dot{E}$ \\
(d)    & (R$_{\odot}$)& (km s$^{-1}$)& (km s$^{-1}$)& (R$^2$/d) & erg s$^{-1}$\\
\hline
 2.5  &  16.11        & 326.26   &   130.00 & 0.0043  & 1.38$\times$10$^{34}$ \\  
 5.0  &  25.58        & 258.96   &    64.80 & 0.0015  & 2.59$\times$10$^{32}$  \\
 7.5  &  33.52        & 226.22   &    43.20 & 0.0024  & 2.20$\times$10$^{31}$  \\
10.0  &  40.60        & 205.53   &    32.40 & 0.0028  & 3.99$\times$10$^{30}$  \\
12.5  &  47.12        & 190.80   &    25.90 & 0.0030  & 1.06$\times$10$^{30}$  \\
15.0  &  53.21        & 179.55   &    21.60 & 0.0031  & 3.45$\times$10$^{29}$   \\
17.5  &  58.97        & 170.55   &    18.50 & 0.0031  & 1.39$\times$10$^{29}$   \\
20.0  &  64.46        & 163.13   &    16.20 & 0.0031  & 6.16$\times$10$^{28}$  \\
22.5  &  69.72        & 156.85   &    14.40 & 0.0031  & 3.02$\times$10$^{28}$  \\
25.0  &  74.80        & 151.44   &    13.00 & 0.0031  & 1.60$\times$10$^{28}$  \\
\hline
\end{tabular}
\end{table}

\subsection{Dependence on $\nu$ and $\delta$R/R$_1$}

The viscosity and the thickness of the oscillating layer are free parameters that
are not {\it a priori} constrained.  We find that for a given orbital separation, 
the energy dissipation rates may span a range of up to 2 orders of magnitude, 
depending on the combination of $\nu$ and $\delta$R/R$_1$.  In Figure 4 we illustrate 
the dependence of the energy dissipation rate on viscosity (left panels) and on the 
thickness of the layer (right panels).  For $P_{orb}=2.5$ days, the energy dissipation rates  
decrease with increasing viscosity.  Since the explicit dependence of $\dot{E}$ on the 
viscosity is linear,  (see equation 17) the energy dissipation would be expected to increase 
for increasing viscosity. On the other hand, because there is also a strong dependence of 
the dissipation rates on the amplitudes of the oscillations, and larger viscosities tend to reduce
the amplitude of oscillation, the net result is a decrease in dissipation rates for
larger viscosities.  We thus see  that for the closest binary systems, it is this decrease 
in oscillation amplitudes with increasing viscosities that dominates the behavior of 
$\dot{E}$.

\begin{figure}
\centering
\includegraphics[width=9cm]{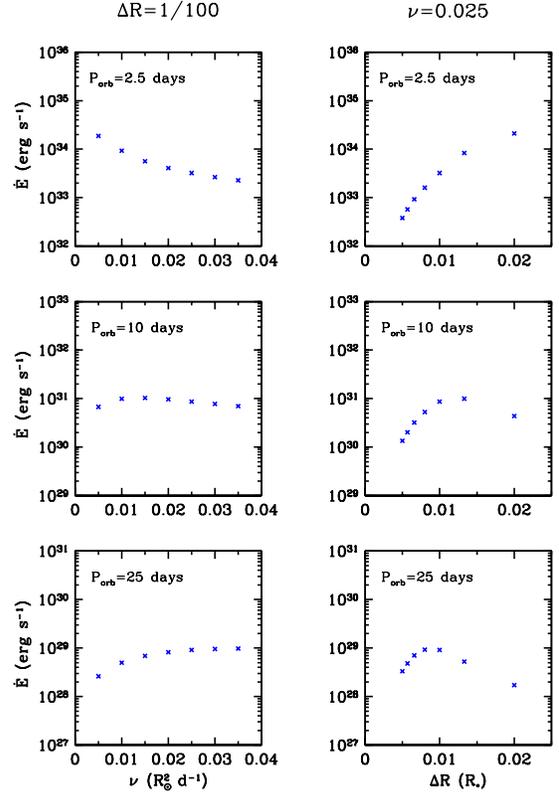}
\caption{Energy dissipation rates calculated for different viscosities $\nu$
with a  fixed layer thickness $\delta$R=0.01 R$_1$ (left column)  and for a
changing layer thickness with a fixed viscosity $\nu=$0.025  R$_\odot^2$ d$^{-1}$ (right column).
Three binary systems with different orbital periods are illustrated, from top to bottom:
P$_{orb}=$2.5, 10 and 25 d.
}
\end{figure}

For the longer orbital periods ($P_{orb}=10$ days), the interplay between the
effect of increasing viscosity and decreasing oscillation amplitudes becomes
more clear.  In the middle panel of Figure 4 we see how $\dot{E}$ first increases
with increasing $\nu$ and then decreases as the oscillation amplitudes rapidly
decline, due to the continued increase in $\nu$.  

The thickness of the layer also has a strong influence on the value
of $\dot{E}$.  In the binary with the shortest orbital period, $\dot{E}$  increases 
systematically by up to 2 orders of magnitude when increasing $\delta R$ 
from R$_1$/200 to R$_1$/50. For longer period binaries, however (middle and
bottom panels, right), there is a critical point at which $\dot{E}$ reaches a
maximum value and then starts to decrease with increasing $\delta$R/R$_1$.  
 For the $P_{orb}=10$ days system this
critical value is $\delta$R$_{c}\sim$1/75 R$_1$ while for the
$P_{orb}=25$ days case the turn around point appears at $\delta$R$_{c}\sim$1/100 R$_1$. 
Hence, the dependence of dissipation on the thickness of the layer
is approximately linear up to the thickness $\delta$R$_{c}$/R$_1$, with $\dot{E}$
increasing monotonically.  However, once the layer is thicker than $\delta$R$_{c}$/R$_1$,
the surface layer has more inertia and its azimuthal velocity amplitudes become
smaller and, once again, it is this effect that dominates the trend in $\dot{E}$.

\section{Discussion}

\subsection{The kinematical viscosity}

The  energy dissipation rates computed from the results of the one-layer stellar oscillation
model for  binary systems  lead to the possibility of constraining the value of the viscosity 
$\nu$. This can be performed by comparing the value of the synchronization timescale for a given 
binary system as obtained from our model $\tau_{sync}$, with the synchronization timescale curve 
predicted by  Zahn's relation for the same system, $\tau_{Zahn}$. The difference between
$\tau_{sync}$ and $\tau_{Zahn}$, if attributed to viscosity alone, allows its value to be determined. 
For the system parameters listed in Table 1 we derive 
$\nu$ in the range 0.0015--0.0043 R$_\odot^2$ d$^{-1}=$0.84--2.41$\times$10$^{14}$ cm$^2$ s$^{-1}$.  
If we apply the analogy with accretion disk studies where the viscosity is incorporated in the 
$\alpha-$parameter, an approximate  relationship between kinematical viscosity and the 
Shakura \& Sunyaev $\alpha$ value is

\begin{equation}
\nu\sim\alpha \frac{kT}{\mu m_H} \frac{1}{\omega}
\end{equation}
\noindent from where
\begin{equation}
\alpha \sim 0.0981 \frac{(\Delta V_\phi/km/s)\mu (\nu/R_\odot^2/day)}{(T/(10000^\circ K)(R_1/R_\odot)}
\end{equation}

\noindent Adopting the results illustrated in Figure 2 for the maximum azimutal velocity, 
$\Delta$V($\phi$),  assuming a mean molecular weight for a fully ionized gas, $\mu=$0.62, 
T$\sim$10000$^\circ$K, and using  $\nu=$0.005 R$_\odot^2$d$^{-1}$,  the above equation yields  
values of $\alpha\leq$0.002.

It is interesting also to compare this result with the values of the corresponding sum of
diffusion coefficients for rotational mixing that are computed by the
Binary Evolutionary Code (BEC; Petrovic et al. 2005, and references therein).
In order to make this comparison, BEC was run for a P$_{orb}=$ 5 d binary system with
masses 5+4 M$_\odot$.  The computation was stopped after an evolutionary time of
$\sim$10$^6$ years in order to examine the values of these coefficients.  The $\delta$R$=$0.01 R$_1$
layer that we adopted for the oscillations calculations is equivalent to 73 layers of the
BEC  computation,  and the sum of the values of the combined diffusion coefficients for rotational 
mixing \footnote{see Heger (1998) for a summary and description of the diffusion coefficients that
are computed} from these layers is 1.65$\times$10$^{14}$ cm$^2$ s$^{-1}$ ($=$0.0030 R$_\odot^2$/d). 
It is not, however, clear how the values obtained for these 73 layers should be combined in order 
to compare the result with the $\nu$ parameter that is used in the one-layer calculation, and the
sum is most likely an overestimate.  On the other hand, this value does not include the effects 
due to a magnetic field, which would tend to increase the value of the diffusion coefficient for
each layer.

\subsection{Evolution of energy dissipation rates}

The energy dissipation rate, which is extracted from
the rotational energy $E_{rot}$ of the star through the action of de-spinning, is calculated from the
difference between its initial rotational energy and the energy at the time of synchronization
divided by the synchronization timescale. Thus, this  represents  an average
of the energy dissipation rate over the time it takes the star to reach synchronization.  However,
the rate of dissipation is expected to be a function of time, gradually vanishing
as the system approaches synchronization.  Indeed, Figure 5 illustrates the dependence on $\beta$ of
the energy dissipation rate due to azimuthal motions, showing that as $\beta\rightarrow$1,
$\dot{E}\rightarrow$0.  Figure 5 corresponds to the model binary system with P$_{orb}=$5 days,
$\nu=$0.005 R$_\odot^2$ day$^{-1}$, and other parameters as used for  Table 1.

\begin{figure}
\centering
\includegraphics[width=9cm]{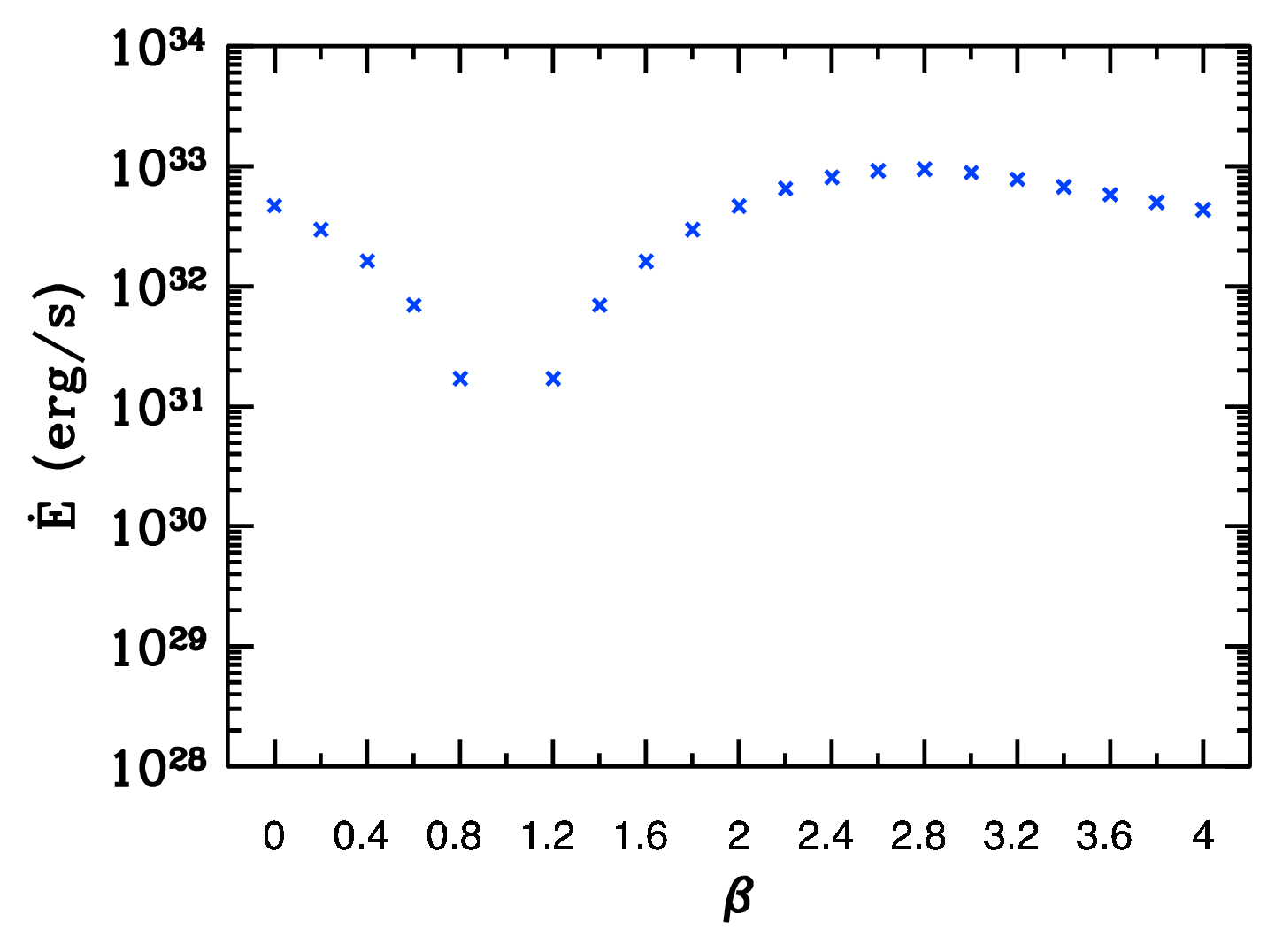}
\caption{Energy dissipation rates for different values of $\beta$, for a binary system with
M$_1=$5 M$_\odot$, M$_2=$4 M$_\odot$, R$_1=$3.2 R$_\odot$, P$_{orb}=$5 d, $\delta$R$=$0.01 R$_1$
and $\nu=$0.005 R$_\odot^2$ d$^{-1}$.
}
\end{figure}

Though we have not addressed the case of excentric binary systems in this paper, it is important to
note that due to the  dependence of angular velocity with orbital separation,
$\beta$ is a function of orbital phase.  This effect, combined with the variation in orbital
separation, leads to strongly varying energy dissipation rates over the orbital cycle.  
An exhaustive study of the available parameter space is beyond the scope of the
present paper, but will be the objective of a forthcoming investigation.
It is, however, worth noting that within the framework of Zahn's (1977) theory,
the synchronization timescales are several orders of magnitude smaller than the
circularization timescales.  Hence, all binary systems should evolve to pseudo-synchronization
(i.e., {\em average} orbital angular velocity$=$ rotational angular velocity) long
before the orbit is circularized.  Thus, large numbers of asynchronous binary systems 
in circular orbits are not expected to exist for stars on the Main Sequence.  

\subsection{Non-synchronous rotation and surface activity}

An interesting consequence suggested by Figure 5 is that non-synchonous binary systems will
tend rapidly towards values of $\beta$ near unity, but will live for a relatively longer time
near the equilibrium value, but without having achieved it.  Given the very large uncertainties
in the observable parameters, primarily V$_{rot}$ {\rm sin i} and R$_1$,  it is not easy to
establish whether a particular binary system has actually {\em achieved} equilibrium.  Thus,
there may be a large number of non-synchronously rotating binary systems in which observable
effects due to the tidal oscillations may be present.  In particular, the peculiar X-ray emission
that is observed in a small number of B-star binary systems in which there is evidence
for non-synchronous rotation (Haro et al. 2003) may be a manifestation of tidally-induced
activity.  Although the detailed mechanism for converting tidal energy into X-ray emission
is not known, we have previously speculated (Moreno et al. 2005; Koenigsberger et al. 2006)  
that the shear produced by the relative azimuthal motions of the outer stellar layers  
may lead to magnetic field generation near the stellar surface as well as  mass ejection
episodes.  Thus, the X-ray emission could be associated with these processes.
The possible connection between surface activity and tidal interactions has already
been investigated in connection with the distribution of star spots in RS CVn-type
binary systems (Holtzwarth \& Sch\"ussler (2002).  

Other observational manifestations of this activity may be ``turbulence" above the expected 
value for the effective temperature of the star,  and photometric and line profile variability.
For example, the theoretical absorption line profiles that were computed for the optical
counterpart of the X-ray binary system 2S0114$+$650 required a ``macroturbulence"
parameter of 37.5 km s$^{-1}$ to match the observations (Koenigsberger et al. 2006),
significantly larger than the thermal speeds expected in a B1-supergiant star. However,
the problem of excessively large ``macroturbulent" velocities is widely encountered
among massive stars and, to our knowledge, there has been no systematic study made to
determine whether the problem is more severe in binary stars than in single stars. 

\section{Conclusions}

We have shown that a single-layer approximation for the calculation of tidal oscillations
yields results that are consistent with the predictions of  Zahn's (1966, 1977, 1989) theory 
for the synchronization timescales in circular orbits, $\tau_{sync}\sim a^6$,
thus providing a simplified  means of computing the energy dissipation rates, $\dot{E}$
in binary systems.  Furthermore,  by calibrating our model results to fit Zahn's relationship, 
we are able to constrain the value of the kinematical viscosity parameter, $\nu$. For the binary
system models considered in this investigation, $\nu\sim$0.0015--0.0043 R$_\odot ^2$/day.  This is
significantly larger than values typically obtained from detailed computations of the interior
structure of stellar convective layers, in the absence of magnetic fields.

Our code makes no  assumption regarding the energy dissipation mechanism except that 
the dissipation processes can be described through  the parameter $\nu$.  We 
show in this paper that if we assume that the energy dissipation mechanism is related to 
turbulent viscosity, we arrive at Equation 19, which allows us to compute  energy
dissipation rates using the output of the code, and we show that the derived energy dissipation
rates are consistent with those predicted by theoretical models of stars with convective envelopes.
However, the code includes no {\it a priori} assumptions regarding the energy transport mechanism.
In order to compare the results of the code with the theory for stars having outer radiative
envelopes,  the effects due to radiative damping need to be incorporated in the derivation
of the synchronization timescale, which goes beyond the scope of the present investigation.

In summary, however, we have shown that the Moreno et al. (2005) one-layer model yields 
results that are consistent with other theoretical calculations of the tidal interactions
in binary systems.  It has the advantage that it can be used to compute the tidal oscillations  
for arbitrary stellar rotational velocities and orbital excentricities,  allowing an estimate 
for the numerical value of $\nu$ under general conditions in  binary systems.  
In addition, the code  provides values of the radial and tangential components of the
oscillation velocities  at any time throughout the orbit, as well as  the location on the
stellar surface where the largest shear instabilities may be occurring. This allows 
a comparison with observational quantities, such as absorption line profiles and diagnostics
of surface activity  and should allow a methodical exploration of the impact that tidal
oscillations may have on the surface properties of stars in non-synchronously rotating
binary systems.


\section{Acknowledgements}

GK thanks S$\phi$ren Meibom, Luis Mochan and Daniel Proga for helpful discussions. Support from 
CONACYT grant 36569 and UNAM/PAPIIT IN119205 are acknowledged.





\begin{thebibliography}{}{}

\bibitem{} Claret, A. \& Cunha, N.C.S. 1997, A\&A, 318, 187.
\bibitem{}Frank, J.,King, A.R., Raine, D.J., 1985, Accretion Power in Astrophysics, Cambridge University Press, ISBN 0521245303, pp.65  
\bibitem{} Gies, D.R., and  Kullavanijaya, A. 1988, ApJ, 326, 813.
\bibitem{} Heger, A. 1998, Ph.D. Dissertation, Technische Universit\"at, M\"unchen, Fakult\"at f|"ur
Physik, MPA 1120.
\bibitem{} Holzwarth, V. \& Sch\"ussler, M., 2002, AN 323, 399
\bibitem{} Hut, P. 1981, A\&A, 99, 126.
\bibitem{} Landau, L.D. and Lifshitz, E.M. 1984, Fluid Mechanics, Pergamon Press.
\bibitem{} Lynden-Bell and  Pringle, 1974, MNRAS, 168, 803.
\bibitem{} Mathews L.D., Mathieu R.D., 1992, ASP conf. ser. 32, 244 (H. McAlister, W.I. Hartkopf, (eds.)
\bibitem{} Moreno, E \& Koenigsberger, G. 1999, RMA\&A, 35 157  
\bibitem{} Moreno, E., Koenigsberger, G. \& Toledano, O. 2005, A\&A, A\&A, 437, 641.  
\bibitem{} Pan, K. 1997, A\&A 321, 597.
\bibitem{} Petrovic, J., Langer, N., Yoon, S.-C., \& Heger, A. 2005, A\&A, 435, 247.
\bibitem{} Spruit, H.C.  2002, A\&A 381, 923
\bibitem{} Shakura, N.I. \& Sunyaev, R.A. 1973, A\&A, 24, 337.
\bibitem{} Smith, M.A. 1977,  ApJ, 215, 574
\bibitem{} Tassoul, J.-L. 2000, Cambridge Unversity Press, ISBN 0521772184 
\bibitem{} Willems, B., \& Aerts, C. 2002, A\&A, 384, 441.
\bibitem{} Witte M.G. \& Savonije G.J., 1999, A\&A, 350, 129
\bibitem{} Yoon, \& Langer, N. 2005, A\&A 443, 643.
\bibitem{} Zahn, J.P. 1966, Annales D'Astrophysique $29^{e}$ Ann\'{e}e No.4   
\bibitem{} Zahn, J.P. 1975, A\&A, 41,329  
\bibitem{} Zahn, J.P. 1977, A\&A, 57,383  
\bibitem{} Zahn, J.P. 1989, A\&A, 220,112  

\end{thebibliography}
\end{document}